# Monolithically integrated active passive waveguide array fabricated on thin film lithium niobate using a single continuous photolithography process

*Yuan Zhou,*[1,2] *Yiran Zhu,*[3] *Zhiwei Fang,*[3,*] *Shupeng Yu,*[1,2] *Ting Huang,*[3] *Junxia Zhou,*[3,4] *Rongbo Wu,*[3] *Jian Liu,*[3] *Yu Ma,*[1,2] *Zhe Wang,*[3] *Jianping Yu,*[1,2] *Zhaoxiang Liu,*[3] *Haisu Zhang,*[3] *Zhenhua Wang,*[3] *Min Wang,*[3] *And Ya Cheng*[1,3,4,5,6,7,*]

*Corresponding Author: E-mail: zwfang@phy.ecnu.edu.cn; ya.cheng@siom.ac.cn

[1]State Key Laboratory of High Field Laser Physics and CAS Center for Excellence in Ultra-intense Laser Science, Shanghai Institute of Optics and Fine Mechanics (SIOM), Chinese Academy of Sciences (CAS), Shanghai 201800, China
[2]Center of Materials Science and Optoelectronics Engineering, University of Chinese Academy of Sciences, Beijing 100049, China
[3]The Extreme Optoelectromechanics Laboratory (XXL), School of Physics and Electronic Science, East China Normal University, Shanghai 200241, China
[4]State Key Laboratory of Precision Spectroscopy, East China Normal University, Shanghai 200062, China
[5]Collaborative Innovation Center of Extreme Optics, Shanxi University, Taiyuan 030006, China
[6]Collaborative Innovation Center of Light Manipulations and Applications, Shandong Normal University, Jinan 250358, People's Republic of China.
[7]Shanghai Research Center for Quantum Sciences, Shanghai 201315, China

**Abstract**:   We demonstrate a robust low-loss optical interface by tiling passive (i.e., without doping of active ions) thin film lithium niobate (TFLN) and active (i.e., doped with rare earth ions) TFLN substrates for monolithic integration of passive/active lithium niobate photonics. The tiled substrates composed of both active and passive areas allow to pattern the mask of the integrated active passive photonic device at once using a single continuous photolithography process. The interface loss of tiled substrate is measured as low as 0.26 dB. Thanks to the stability provided by this approach, a four-channel waveguide amplifier is realized in a straightforward manner, which shows a net gain of ~5 dB at 1550-nm wavelength and that of ~8 dB at 1530-nm wavelength for each channel. The robust low-loss optical interface for passive/active photonic integration will facilitate large-scale high performance photonic devices which require on-chip light sources and amplifiers.

# 1. Introduction

Single crystalline lithium niobate (LN) is an attractive photonic material owing to its wide transparent window, moderately high refractive index, as well as large acousto-optic, nonlinear and electro-optic coefficients. Turning the bulk LN crystal into thin film lithium niobate (TFLN) has further enabled the fabrication of high-performance integrated photonic devices for both classical and quantum applications, such as low-loss waveguides, high-quality microresonators, high-speed modulators and high-efficiency optical frequency converters, etc.[1-4] To realize integrated active photonic devices, TFLN doped with rare earth ions (REI) has recently been employed to demonstrate micro-lasers, waveguide amplifiers, quantum emitters and quantum memories.[5-21] Nevertheless, monolithic integration of the active devices fabricated on the REI-doped TFLN with passive TFLN photonic devices has not been demonstrated due to the challenging difficulties in achieving the high precision alignment, low-loss interfacing, and reliable bonding. To this end, the traditional strategy of active passive integration requires use of additional coupling elements, such as on-chip spot size converter (SSC) and lensed fiber, which leads to significant increase of complexity and cost of manufacturing and degradation of performance of the integrated devices.

In this work, we demonstrate for the first time to the best of our knowledge, a robust low-loss optical interface for passive and active lithium niobate photonics by tiling the commercially available TFLN with an REI-doped TFLN substrate before the lithographic fabrication process. Afterwards, a single continuous photolithography process is conducted for patterning the mask of the integrated devices followed by a chemo-mechanical etching for transferring the mask pattern to the TFLN substrate. The fabrication technique, which is coined photolithography assisted chemo-mechanical etching (PLACE), has enabled fabrication of large-scale photonic devices of low propagation loss. [22] Here, we demonstrate an optical interface of active passive photonic integration with an insertion loss of 0.26 dB, which has been used to produce a four-channel waveguide amplifier. The optical gain performance has been characterized in each channel of the waveguide amplifier, featuring a net gain of ~5 dB at 1550-nm wavelength and that of ~8 dB at 1530-nm wavelength for each channel. The device provides convincing evidence that the developed approach is of practical use for a wide range of photonic applications which require monolithic integration and low-loss interfacing of active and passive photonic devices.

## 2. Device Fabrication

**Figure 1** depicts the schematic process flow of the optical interface fabrication for the integrated passive and active TFLN photonics. Firstly, we prepare both the undoped TFLN On Insulator (TFLNOI: 500-nm TFLN/2-µm $SiO_2$/500-µm Si) substrate and the REI doped TFLNOI substrate [**Figure 1**(a)] which will be further used in the tiling process. Before we conduct the tiling, both the doped and undoped TFLNOI substrates are coated with a 200-nm-thick chromium (Cr) film as a hard mask material for subsequent chemical mechanical polishing (CMP) process. The sidewalls of the two TFLNOI substrates around the optical interface are polished into smooth and highly vertical surfaces which is vital for achieving the low-loss interfacing. Secondly, we tile the undoped and REI doped TFLNOI substrates seamlessly using a home-built fixture tool. The home-built fixture is a simple tool which have two clamps to press the two TFLNOI substrates for reducing the air gap in between [**Figure 1**(c)]. In this step, the undoped and REI doped TFLNOI substrates are flip bonded on a polished glass plate with high flatness for the optimal alignment of the top surfaces of two TFLNOI substrates [**Figure 1**(b)]. Ultraviolet (UV) glue is applied on the bottom of tilted substrates and a quartz plate is used to support the tiled TFLNOI substrate before the irradiation with UV light [**Figure 1**(c)], and afterwards a laser welding process is conducted to permanently fix the TFLNOI substrates on the quartz support with sufficient mechanical stability and rigidity [**Figure 1**(d)]. Thirdly, the optical waveguide patterns are produced by femtosecond laser selective direct-writing, and the following CMP process is used to transfer the waveguide pattern into the tiled TFLN [**Figure 1**(e) and (f)]. More fabrication details of the PLACE technique can be found in our previous work.[10, 22]

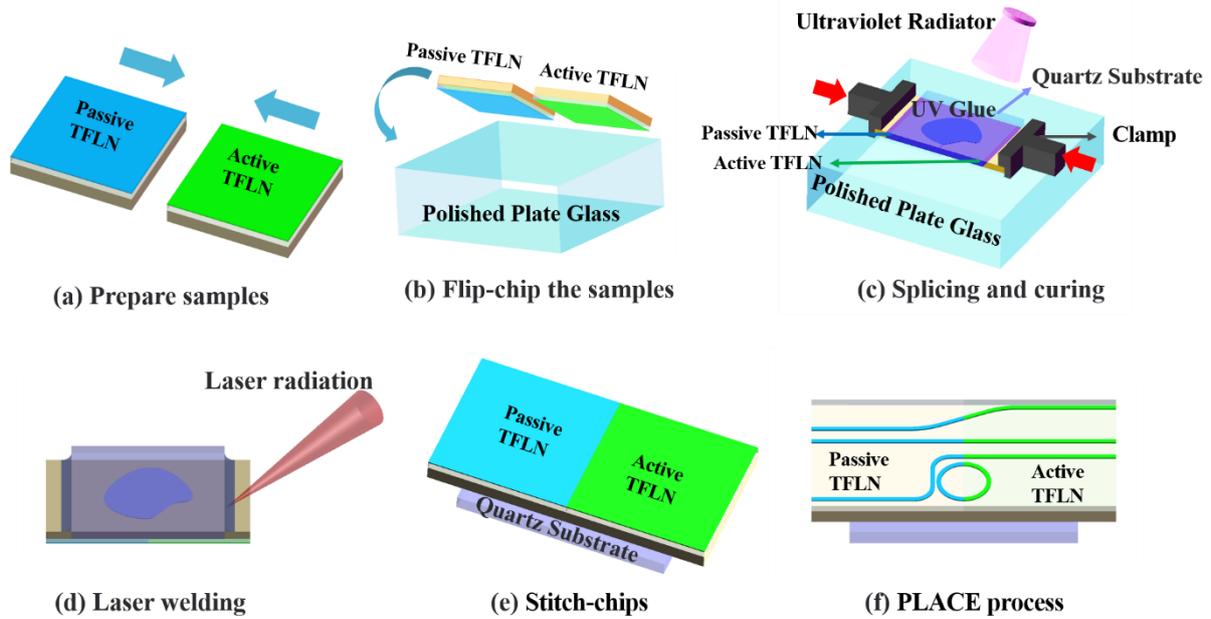

**Figure 1.** Schematic of the fabrication process for robust low-loss optical interface of passive and active lithium niobate photonics. (a) Prepare none-doped and REI doped TFLNOI substrates. (b) None-doped and REI doped TFLNOI substrates are flip-chip on a polished plate glass. (c) The none-doped and REI doped TFLNOI substrates are tiled seamlessly using customized fixture and the UV glue is applied on bottom of stitched chips to fasten the two substrates preliminarily. (d) The followed laser welding is operated in the boundary of two TFLNOI substrates and quartz substrate to achieve a durable bonding. (e)The finished tiled of passive and active TFLNOI. (f)The monolithic passive and active TFLN photonic structures fabricated by PLACE process.

**Figure** 2(a) shows a typical straight TFLN waveguide fabricated on a 500-nm-thick undoped TFLN substrate. **Figure** 2(b) and (c) show integrated active passive TFLN waveguides with perpendicular and tilted interfaces. Both the doped and undoped TFLN substrates have a thickness of 500 nm before they are tiled into the integrated substrate. The angled physical contact (APC) connector in **Figure** 2(c) differs from the physical contact (PC) connector in **Figure** 2(b) for that it can relieve the back reflection with the Brewster angle chosen at the interface. The PLACE process naturally forms a slope on the waveguide sidewall, resulting in ridge-shaped cross section. The detail can be found from our previous publication.[23] The schematic of the waveguide is shown in **Figure 2**(d), where $w_0 = 1\ \mu m$, $w_1 = 1.14\ \mu m$, $w_2 = 4.8\ \mu m$, $T_0 = 90\ nm$, $T_1 = 120\ nm$, $T_2 = 290\ nm$. **Figure 2**(e) is a scanning electron microscopy (SEM) image of the cross section of a TFLN waveguide fabricated by PLACE process. As shown in **Figure 2**(f), the integrated TFLN waveguide is also characterized by SEM, and the gap at the optical interface is determined to be less than 22 nm.

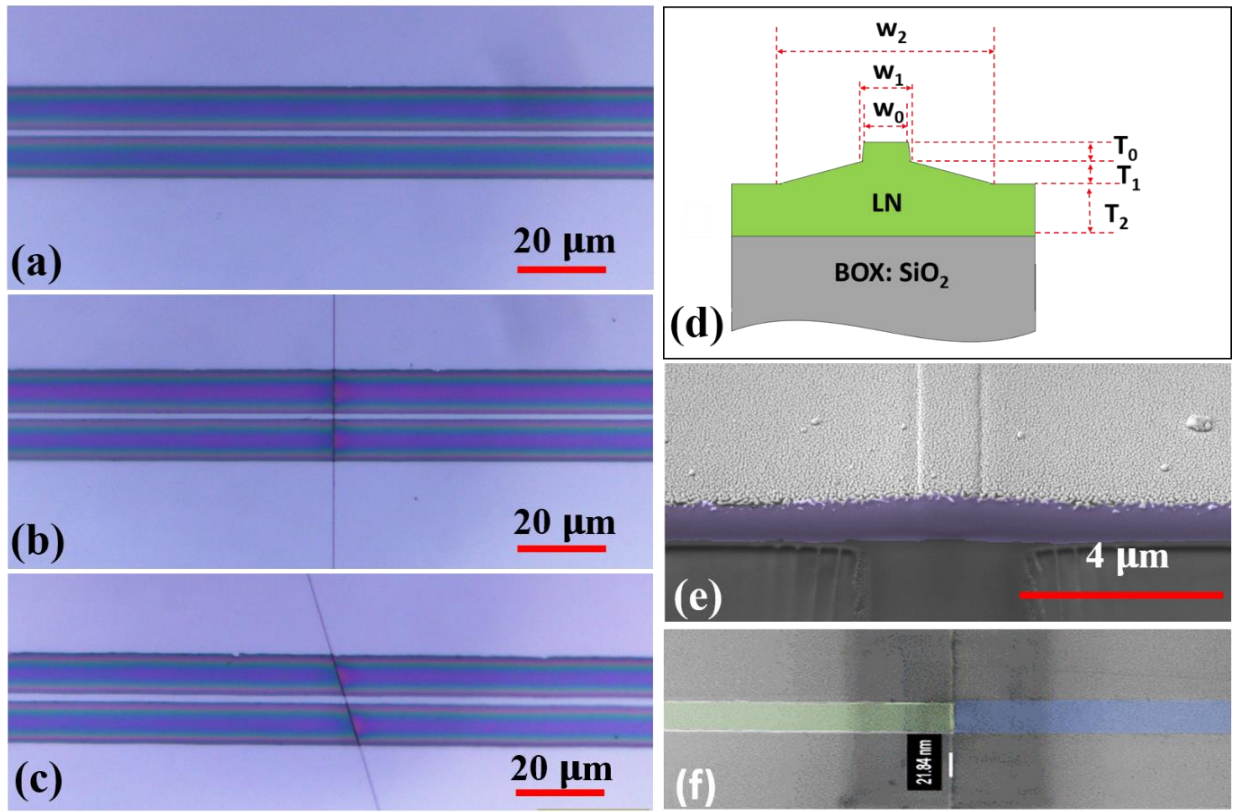

**Figure 2**. Optical micrographs of the fabricated (a) straight monolithic none-doped TFLN waveguide, (b) straight seamless-stitching TFLN waveguide, and (c) oblique seamless-stitching TFLN waveguide. (d) Schematic of ridge waveguide structure on TFLN fabricated by PLACE technique. (e) Scanning electron microscopic (SEM) image of the cross section of a TFLN waveguide fabricated by PLACE technique. (f) Top view SEM image of straight seamless-stitching TFLN waveguide.

## 3. Results and Discussions

We design a structure as shown in **Figure** 3(a) to measure the coupling loss between the active and passive segmemnts of the integrated waveguide which is mainly caused by the slight misalignment in the vertical direction between the passive and active TFLN substrates. A beam splitter based on a multimode interference (MMI) coupler which can realize 50/50 beam splitting at 1550-nm wavelength divides the beam into two beams of equal power. One waveguide arm with multiple bends with a bending radius of 300 nm passes through the interface five times, while the other waveguide arm of the same geometry only passes the interface once. We send the 1550-nm wavelength laser beam into the beam splitter and the output beam is collected by an objective lens and imaged with an infrared camera. By comparing the beam powers measured from the two output ports, the loss caused by the extra four passings through the interface can be calculated by

$$\alpha = -10 lg \sqrt[4]{\frac{P_{out1}}{P_{out2}}}$$

where $P_{out1}$ is the output power of the waveguide arm passing through the interface five times, and $P_{out2}$ is the power of the output of the waveguide arm passing through the interface only once. For this sample, we determine that the height difference is around 60 nm. To minimize the measurement error, we repeat the measurement eight times at the differnt sinput powers and record the output powers to generate eight sets of data. The average insertion loss of the data set is 0.26 dB, and the standard deviation is 0.044 dB as showed in **Figure 3**(b). **Figure 3**(b) also shows the simulation of the straight integrated active passive TFLN waveguides. We use the software Mode Solution in our simulation to find out the relationship between the loss and height-difference caused misalignment, and the simulation result agrees well with the experiment. It can be clearly seen that the coupling loss at the interface increases with the height difference between the passive and active waveguides. It is very encouraging from the simulation result that the interfacing loss will be less than 1 dB when the height difference is less than 160 nm, which is not difficult to ensure in today's photonic industry.

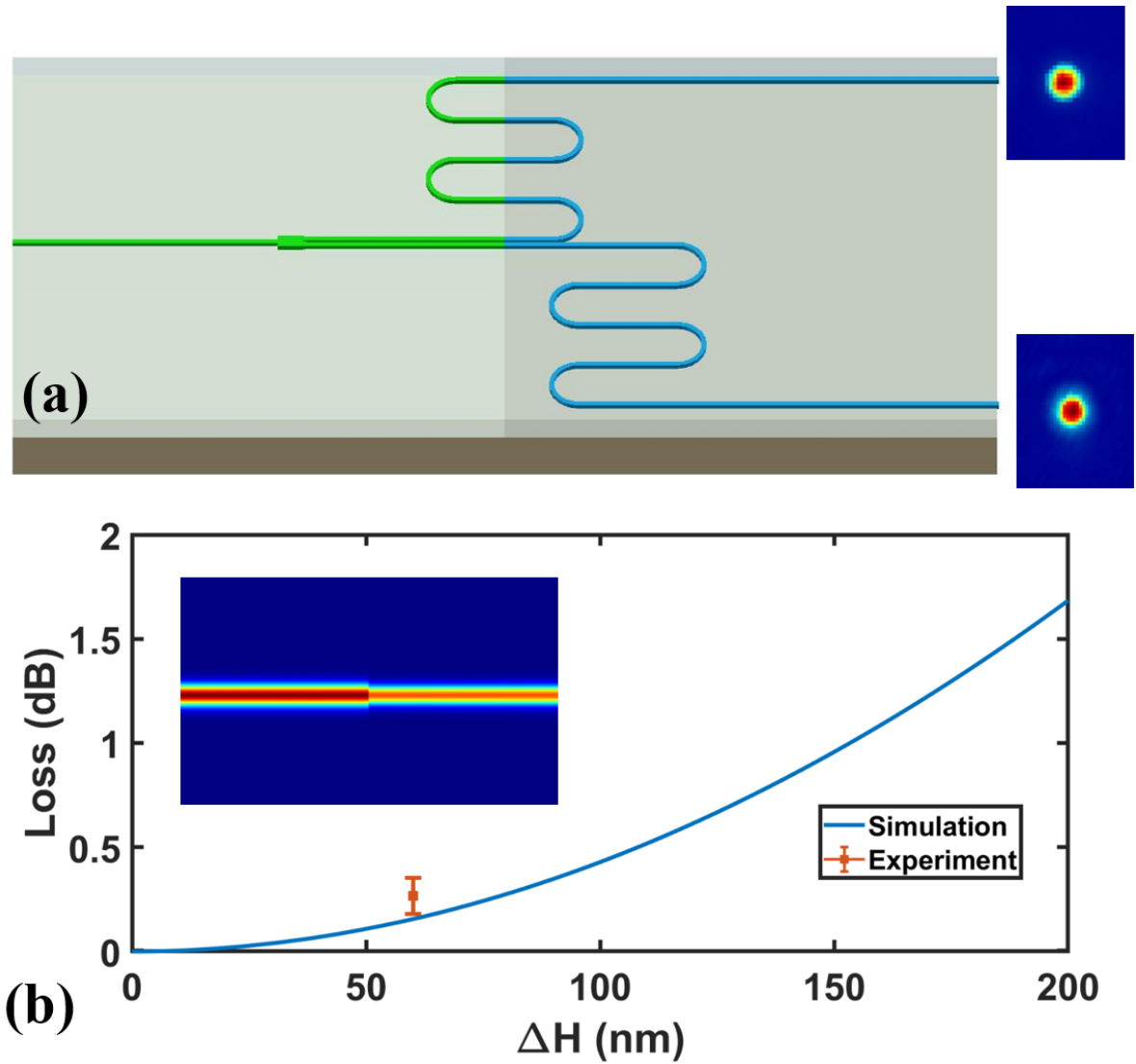

**Figure 3.** (a) A beam splitter based on MMI coupler to measure the insertion loss. (b) Measured (scatters) and simulated (curves) loss for different height differences. The change of light field near the seams as ΔH = 60 nm is shown in the inset.

The design of the four-channel waveguide amplifier is illustrated in **Figure 4**(a), which consists of four $Er^{3+}$-doped spiral waveguides connected by an MMI beam splitter. The four waveguide amplifiers are fabricated on $Er^{3+}$-doped TFLN while the three MMI couplers are fabricated on the undoped TFLN. **Figure 4** (b) shows the digital picture of the integrated device, the length of each $Er^{3+}$-doped waveguide is ~2 cm. The output beam profile of the waveguides is also captured by an objective and imaged onto an infrared camera. As shown in the false-color insets of **Figure 4**(c), the 1550-nm wavelength beams in the four $Er^{3+}$-doped waveguides are all in the fundamental mode with a uniform intensity distribution.

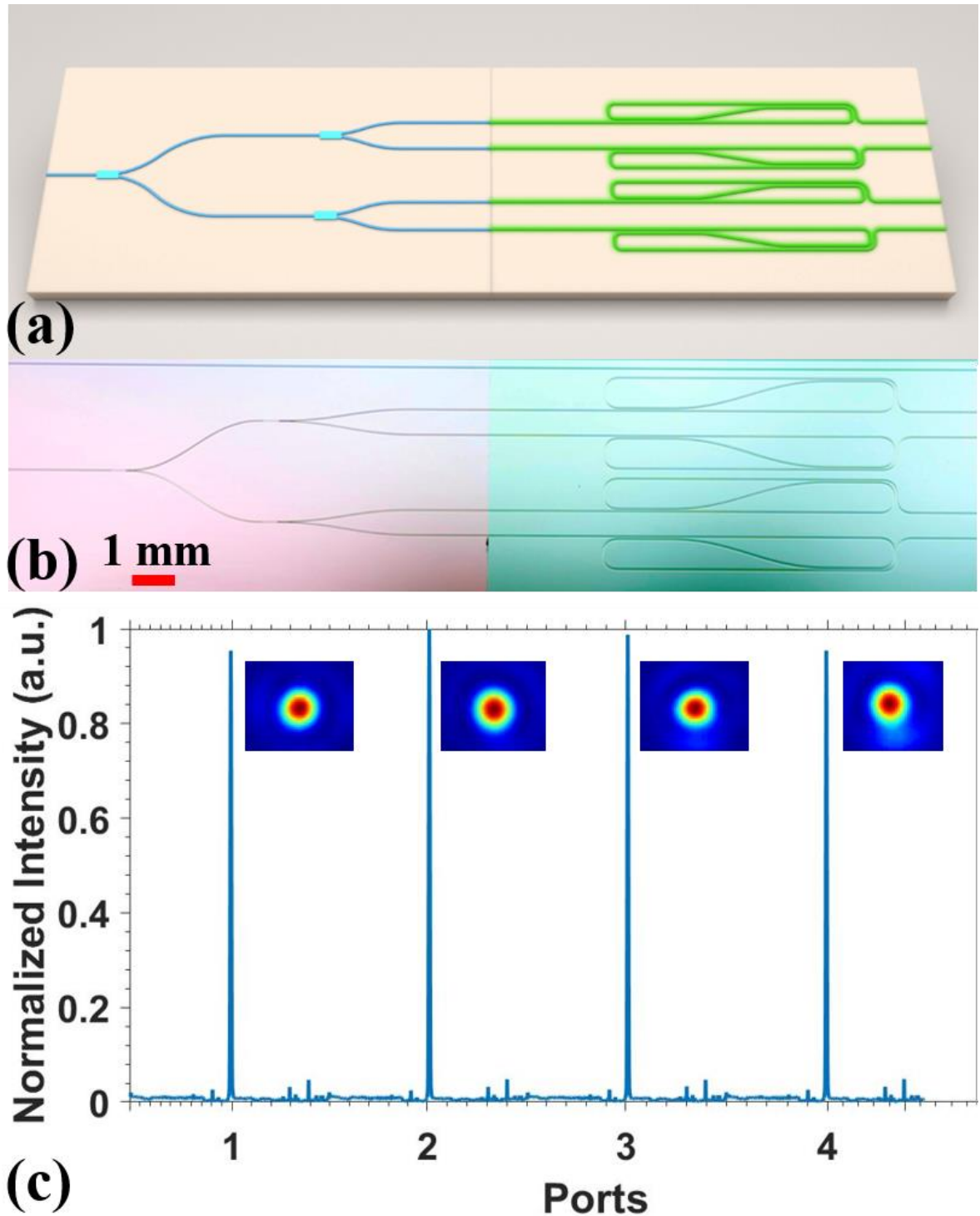

**Figure 4** (a) Illustration of four-channel waveguide amplifiers. (b) Photo of four-channel waveguide amplifiers. (c) The mode and intensity distribution of 1550-nm wavelength signal in the four channel $Er^{3+}$-doped waveguides..

The photograph of the $Er^{3+}$-doped TFLN waveguide amplifier array under the pumping with a 976 nm diode laser is shown in **Figure 5**(a), demonstrating the strong green upconversion fluorescence along the four spiral waveguides. **Figure 5** (b) and (c) demonstrate the net gain of the integrated amplifier as a function of the launched pump power at the signal wavelengths at 1550 nm and 1530 nm, respectively. In both cases a rapid incarese of the gain with the increasing pump power is observed, which is followed by a slow gain saturation at the higher pump power. Specifically, the maximum internal net gain for the 1550 nm signal reaches ~5 dB and that for the 1530 nm signal reaches ~8 dB, which is quite uniform in the four waveguide amplifiers. This shows the stability in the fabrication process as the four waveguide amplifiers are designed to have the same parameters. This multi-channel waveguide amplifier is expected to contribute to the high-power output from on-chip amplifiers by coherent beam combination.

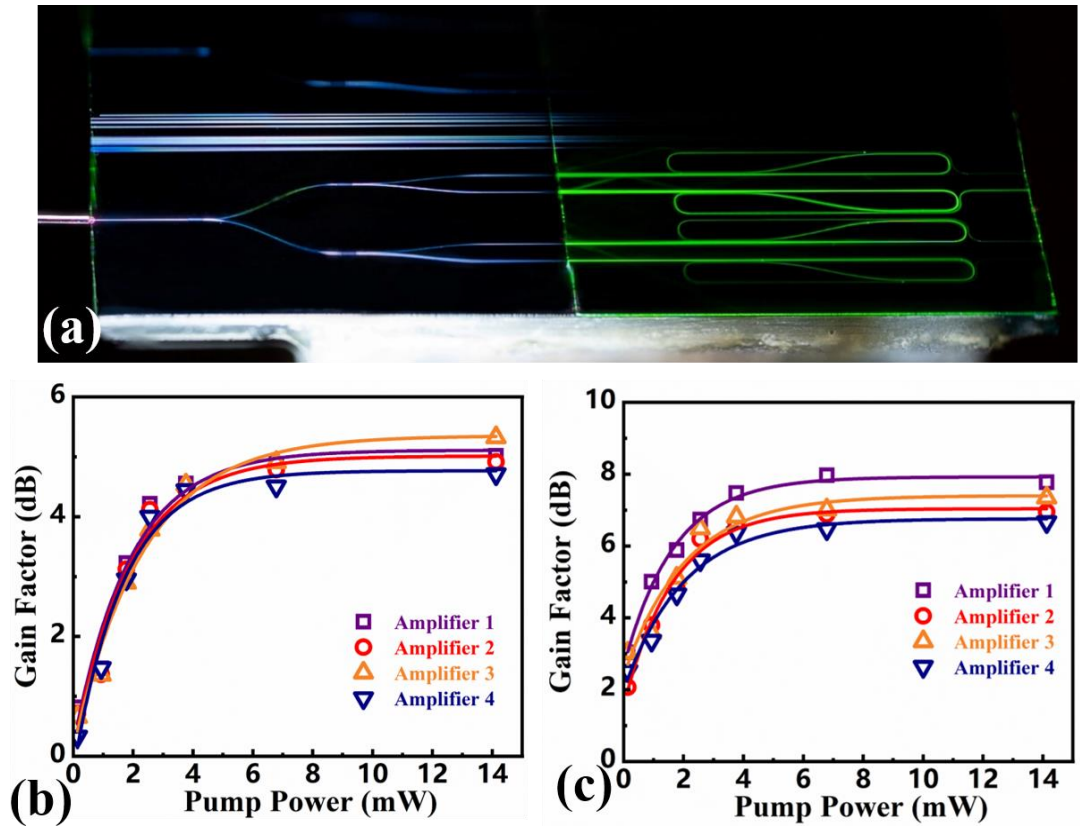

**Figure 5.** (a) Photo of four-channel waveguide amplifiers array when pumped by 976-nm diode laser. Gain characterization of the four $Er^{3+}$-doped LN waveguides array for signal wavelengths of (b) 1550 nm and (c) 1530 nm.

## 4. Conclusion

In summary, we demonstrate a robust low-loss optical interface for monolithic integration of passive and active TFLN photonics by tiling the undoped and REI doped TFLN substrates followed with a single continuous photolithography fabrication process. We achieve a coupling loss of 0.26 dB at the waveguide interface. Furthermore, a four-channel waveguide amplifier array is fabricated, showing a net gain of ~5 dB at 1550-nm wavelength and ~8 dB at 1530-nm wavelength in each $Er^{3+}$-doped waveguide, respectively. The strategy of tiling passive and active TFLN substrates to form low-loss optical interfaces for the monolithic integration of passive and active TFLN photonics offers advantages in its high scalability, high reliability, high production rate and cost-effectiveness.